\documentclass[preprint,aps,pra,showpacs,floatfix,superscriptaddress]{revtex4-1}
\usepackage{longtable}
\usepackage[referable]{threeparttablex}

\usepackage{graphicx}
\usepackage{amsmath,amsfonts,amssymb,verbatim,float}

\usepackage[usenames]{color}	
\usepackage{color,soul}	
\usepackage{colortbl}
\usepackage{calrsfs}
\usepackage{ulem} 

\usepackage{hyperref} 
\usepackage{subfigure}
\usepackage{indentfirst}
\usepackage{siunitx}
\definecolor{faded}{gray}{0.45}
\newcommand{\bp}{{\bf p}}

\newcommand{\br}{{\bf r}}

\newcommand{\hbr}{\hat{{\bf r}}}

\newcommand{\balpha}{\boldsymbol{\alpha}}

\begin{document}

\thispagestyle{empty}
\title{
Complex scaled relativistic configuration-interaction study of the $LL$~resonances in helium-like ions: from Boron to Argon 
}
\author{V.~A.~Zaytsev}
\affiliation{
Department of Physics, St. Petersburg State University,
Universitetskaya naberezhnaya 7/9, 199034 St. Petersburg, Russia
}
\author{I.~A.~Maltsev}
\affiliation{
Department of Physics, St. Petersburg State University,
Universitetskaya naberezhnaya 7/9, 199034 St. Petersburg, Russia
}
\author{I.~I.~Tupitsyn}
\affiliation{
Department of Physics, St. Petersburg State University,
Universitetskaya naberezhnaya 7/9, 199034 St. Petersburg, Russia
}
\author{V.~M.~Shabaev}
\affiliation{
Department of Physics, St. Petersburg State University,
Universitetskaya naberezhnaya 7/9, 199034 St. Petersburg, Russia
}
%
\begin{abstract}
Energies and Auger widths of the $LL$ resonances in He-like ions from boron to argon are evaluated by means of a complex scaled configuration-interaction approach within the framework of the Dirac-Coulomb-Breit Hamiltonian.
The nuclear recoil and QED corrections are also taken into account.
The obtained results are compared with other calculations based on the complex scaling method as well as with the related results evaluated using the stabilization and basis balancing methods.
\end{abstract}
%
%
\maketitle
%
%
%
%
%
\section{Introduction}
Autoionizing states of atomic or ionic systems are the excited states which can decay due to the electron-electron interactions via emission of one (or more) electrons. 
A special place among such states is held by the levels of the $LL$ resonance groups of He-like systems.
The simplicity of these systems makes them attractive for both theoretical and experimental investigations.
The investigations aimed at determining of the energies of these levels are of particular interest for plasma diagnostics~\cite{Beiersdorfer_AJ409_846:1993, Widmann_RSI66_761:1995, Kunze, Tallents}, cosmological~\cite{Porquet_SSR157_103:2010} and fusion research (see, e.g., the review~\cite{Fabian_1991}).
A new interest in studying the characteristics of the $LL$ resonances was caused by the recent experiment~\cite{Muller_PRA98_033416:2018}.
In this experiment, a new level of accuracy for the energy of the autoionizing states of the He-like carbon ion was reached.
Experimental data of such accuracy being complemented by the theoretical predictions of the same precision allow one to set these states as energy-reference standards at synchrotron radiation facilities.
The precise theoretical predictions for the energies of the $LL$ resonances are, therefore, highly demanded.
%
%
\\ \indent
%
%
For the accurate evaluation of the energies of autoionizing states, which are strongly affected by electron correlations, the high-precision many-electron methods such as coupled-cluster and configuration-interaction are required.
These methods, being successfully applied for the calculations of bound-level energies, however, fail when naively applied for the description of resonances.
The energies of such resonances show a strong dependence on the parameters of the basis set, e.g., the convergence of the resonance energy with respect to the number of basis functions, which is one of the basis-set parameters, is very weak or even absent. 
This is explained by the fact that the autoionizing states are embedded into the positive-energy continuum.
As a result, they can not be described by square-integrable functions which form the basis set of the coupled-cluster and configuration-interaction methods.
This problem can be naturally solved with the usage of the complex scaling approach which is based on the analytical properties of the spectrum of a Hamiltonian being dilated into the complex plane.
The first mathematical analysis of these properties was performed in Refs.~\cite{Balslev_CMP22_280:1971, Aguilar_CMP22_269:1971} for the nonrelativistic Hamiltonian and in Refs.~\cite{Weder_JMP15_20:1974, Weder_JFA20_319:1975, Seba_LMP16_51:1988} for the relativistic one.
In these works, it was shown that in the spectra of the dilated Hamiltonian the autoionizing states are separated from the continuum.
The wave functions of these states, therefore, become square-integrable and can be investigated with conventional many-electron methods.
That makes the complex scaling approach a powerful tool for studying for studying properties of resonances appearing in various systems and processes.
As examples, the resonances of nuclei~\cite{Kruppa_PRL79_2217:1997, Aoyama_PTP116_1:2006, Arai_PRC74_064311:2006, Guo_CPC181_550:2010}, few-electron systems~%
\cite{Lindroth_PRA49_4473:1994, Lindroth_PRA52_2737:1995, Themelis_JPB28_L379:1995, Brandefelt_PRA, Pestka_JPB, Bylicki_PRA77_044501:2008,Zhang_PRA85_032515:2012}, and molecules~\cite{Medikeri_1995, Mahalakshmi_JCP115_4549:2001} were investigated with the usage of this method.
More applications, as well as the details of the complex scaling approach, can be found in the reviews~\cite{Reinhardt_ARPC33_223:1982, Junker_AAMP18_107:1982, Ho_PRC99_1:1983, Moiseyev_PR302_211:1998, Lindroth_AQC63_247:2012, Simon_IJQC14_529:1978}.
It is also worth noting that the dilated Hamiltonian is not hermitian but symmetric operator with complex eigenvalues.
The real and imaginary parts of the eigenvalues corresponding to the autoionizing states give the energies and Auger widths of the states, respectively.
%
%
\\ \indent
%
%
Apart from the complex scaling approach, one can apply the stabilization or basis balancing methods.
The stabilization method (SM) was pioneered by Hoil{\o}ien and Midtal~\cite{Holoien_Midtal} and was utilized in numerous investigations~\cite{Isaacson_CPL110_130:1984, Moccia_JPB20_1423:1987, Lefebvre_JPB21_L709:1988, Froelich_JCP91_1702:1989, Mandelshtam_PRL70_1932:1993}.
The basis balancing method (BBM) was worked out by Yerokhin with co-authors just recently~\cite{Yerokhin_PRA96_042505:2017} and was applied for the calculation of the energies of the autoionizing levels of Li-like ions in a wide range of the nucleus charge number~\cite{Yerokhin_JPCRD47_023105:2018}.
Both methods are applied to the conventional hermitian Hamiltonian and, as a result, only the real arithmetic is involved that provides a considerable computational advantage. 
However, the energy of the autoionizing state obtained within SM or BBM can differ from the exact one by a shift arising due to the inappropriate treatment of the interaction with the continuum. 
The advantages of these methods over the complex scaling approach, thus, can be completely lost in some cases.
In view of the considerable progress in experimental accuracy for the energies of the autoionizing states~\cite{Muller_PRA98_033416:2018}, the revision of the applicability of the SM and BBM is required.
%
%
\\ \indent
%
%
In the present paper, we apply the configuration interaction~(CI) coupled with the complex scaling~(CS) approach to solve the Dirac-Coulomb-Breit (DCB) equation for $LL$ resonances of He-like ions in the range from boron to argon.
The configuration space is spanned on the one-electron Dirac orbitals being constructed from the B-splines.
The DCB energies are supplemented with the quantum electrodynamics~(QED), nuclear recoil, and frequency-dependent Breit corrections. 
We also estimate the difference of the energies obtained within the SM and BBM with ones calculated employing the complex scaling approach. 
In case of the $2s^2$ level of the He-like carbon ion, it is found that the energy difference between these three methods exceeds the uncertainty reached in the recent experiment~\cite{Muller_PRA98_033416:2018}.
%
%
\\ \indent
%
%
Units $m_e = \hbar = 1$ and the Heaviside charge unit ($e^2 = 4\pi\alpha$) are used in the paper.
%
%
%
\section{BASIC FORMALISM}
%
%
We start with the formulation of the basic principles of the configuration-interaction with complex scaling approach for the solution of the few-electron DCB equation (for the detailed description see, e.g., the review~\cite{Lindroth_AQC63_247:2012}).
Here we considere the simplest variant of the CS, namely, the uniform complex 
rotation.
In this case, the radial variable $r$ is transformed as 
\begin{equation}
r \rightarrow re^{i\theta}, 
\label{eq:ucr}
\end{equation}
with $\theta$ being a constant rotation angle.
This transformation leads to the following complex rotated DCB Hamiltonian
\begin{equation}
H_{\rm DCB}^{(\theta)} = \sum_j h_{D}^{(\theta)}(j) 
+ e^{-i\theta} \sum_{j<k}\left[V_{C}(j,k) + V_{B}(j,k)\right],
\quad
j,k = 1, \dots, N.
\label{eq:DCB}
\end{equation}
Here $N$ stands for the total number of the electrons and $h_{D}^{(\theta)}$ is 
the scaled one-electron Dirac Hamiltonian given by 
\begin{equation}
h^{(\theta)}_{D}(j) = e^{-i\theta}c\balpha_j\cdot\bp_j
+ (\beta - 1)m_ec^2 + V_{\rm nuc}(r_j e^{i\theta}),
\label{eq:1e_dirac}
\end{equation}
with $\balpha$ and $\beta$ being the Dirac matrices, $\bp$ is the momentum operator, and $V_{\rm nuc}$ is the nucleus potential.
In the present paper, we use the spherical model of the nucleus which is transformed in accordance with the rule~\eqref{eq:ucr},
\begin{equation}
V_{\rm nuc}(re^{i\theta}) = 
\left\lbrace
\begin{aligned}
& -\frac{\alpha Z c}{2R_{\rm nuc}}
\left(3 - e^{2i\theta}\frac{r^2}{R_{\rm nuc}^2}\right), 
& r < R_{\rm nuc}
\\
& -e^{-i\theta}\frac{\alpha Z c}{r}, 
& r > R_{\rm nuc}
\end{aligned}
\right.
\end{equation}
In accordance with Eq.~\eqref{eq:DCB} the Coulomb and Breit interelectronic-interaction operators are given by 
\begin{equation}
V_{C}(j,k) = \frac{\alpha c}{r_{jk}},
\label{eq:ii_coul}
\end{equation}
\begin{eqnarray}
V_{B}(j,k) & = &
\alpha c \left\lbrace
\frac{e^{2i\theta}}{2c^2}
\left[h^{(\theta)}_{D}(j),\left[h^{(\theta)}_{D}(k),r_{jk}\right]\right] 
- \frac{\alpha_j\cdot\alpha_k}{r_{jk}}
\right\rbrace
\\ & = &
-\frac{\alpha c}{2r_{jk}} 
\left[\balpha_j\cdot\balpha_k
+ 
\left(\balpha_j\cdot\hbr_{jk}\right)
\left(\balpha_k\cdot\hbr_{jk}\right)
\right],
\label{eq:ii_breit}
\end{eqnarray}
respectively. 
In Eqs.~\eqref{eq:ii_coul} and~\eqref{eq:ii_breit},
$\hbr_{jk} = \br_{jk} / r_{jk}$ with $\br_{jk} = \br_j - \br_k$ and 
$r_{jk} = |\br_{jk}|$.
Having performed the complex rotation of the DCB Hamiltonian~\eqref{eq:DCB}
we now proceed to the construction of its eigenfunctions.
%
%
\\ \indent
%
%
As in the conventional CI method~\cite{Tupitsyn_PRA68_022511:2003, Chen_PRA52_266:1995}, 
the $N$-electron eigenfunction $\Psi(PJM)$ with the parity $P$, total angular 
momentum $J$, and its projection $M$ is expressed as a linear superposition of the 
configuration-state functions (CSFs) $\Phi(\gamma_r PJM)$
\begin{equation}
\Psi(PJM) = \sum_{r = 1}^{N_{\rm CSF}}c_r \Phi(\gamma_r PJM),
\end{equation}
where $\gamma_r$ stands for all additional quantum numbers which determine uniquely 
the CSF. 
The CSFs are eigenstates of the total angular momentum operators $J^2$ and $J_z$, 
constructed from antisymmetrized products of one-electron Dirac orbitals. 
Here these orbitals are chosen to be the solutions of the scaled one-electron Dirac 
Hamiltonian~\eqref{eq:1e_dirac} of the form
\begin{equation}
\nonumber
\psi^{(\theta)}_{\kappa m}(\br) = \frac{e^{-i\theta}}{r}
\begin{pmatrix}
G^{(\theta)}_\kappa(r) \Omega_{\kappa m}(\hat{\br})
\\
i F^{(\theta)}_\kappa(r) \Omega_{-\kappa m}(\hat{\br})
\end{pmatrix},
\end{equation}
where $\kappa = (-1)^{l + j + 1/2}(j + 1/2)$ is the Dirac quantum number determined by the
angular momentum $j$ and the parity $l$ and $\Omega_{\kappa m}$ is the  
spinor spherical harmonic~\cite{Varshalovich}.
As usual in accordance with the basic principles of the relativistic theory with 
the DCB approximation, the CSF are constructed only from 
positive-energy one-electron Dirac orbitals. 
%
%
\\ \indent
%
%
As already mentioned, autoionizing levels after the complex scaling are described 
by the square-integrable and localized wave functions.
To good accuracy these wave functions can be represented by the 
corresponding solutions of the scaled DCB equation in a spherical cavity of a 
finite radius.
In the present paper, this equation is solved using the dual-kinetic-balance finite basis set method~\cite{Shabaev_PRL93_130405:2004} with the basis functions constructed from B-splines~\cite{Sapirstein_JPB29_5213:1996, Bachau_RPP64_1815:2001},
\begin{equation}
\nonumber
\begin{pmatrix}
G_\kappa^{(\theta)}(r) \\ F_\kappa^{(\theta)}(r)
\end{pmatrix}
=
\sum_{i = 1}^N C_{\kappa, i}^{(\theta)} 
\begin{pmatrix}
B_i(r) \\ \frac{e^{-i\theta}}{2Mc}\left[\frac{d}{dr} + \frac{\kappa}{r}\right]B_i(r)
\end{pmatrix}
+
\sum_{i = N+1}^{2N} C_{\kappa, i}^{(\theta)} 
\begin{pmatrix}
\frac{e^{-i\theta}}{2Mc}\left[\frac{d}{dr} - \frac{\kappa}{r}\right]B_{i-N}(r) \\ B_{i-N}(r)
\end{pmatrix}.
\end{equation}
%
%
%
%
\section{RESULTS AND DISCUSSIONS}
\subsection{
Comparison of the stabilization and basis balancing methods with the complex scaling approach
}
\begin{figure}[h!]
\includegraphics[width=0.75\textwidth]{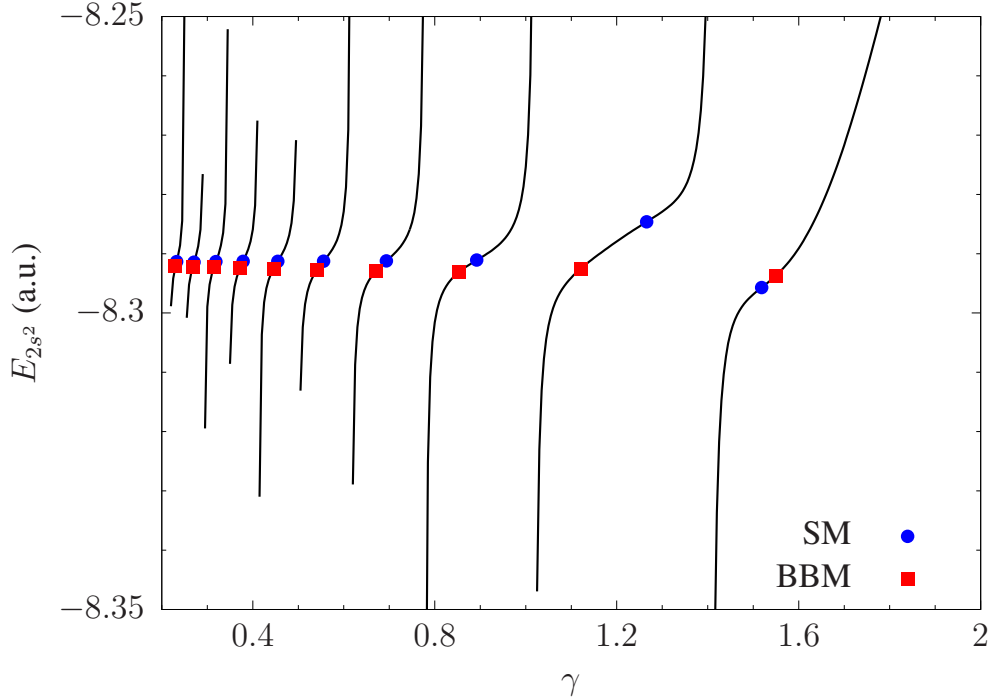}
\caption{
Energy of the $2s^2$ state of the He-like carbon ($Z = 6$) ion as a function of the parameter $\gamma$ (see Eq.~\eqref{eq:grid}).
The CSFs are constructed from one-electron  $s$ and $p$ Dirac orbitals obtained in the basis of $30$ B-splines.
The size of the spherical box was chosen to be 15 a.u.
Blue circles and red squares correspond to the $\gamma$ parameters chosen in accordance with the stabilization and basis balancing methods, respectively.
}
\label{fig:stabilization}
\end{figure}

\begin{figure}[h!]
\includegraphics[width=0.75\textwidth]{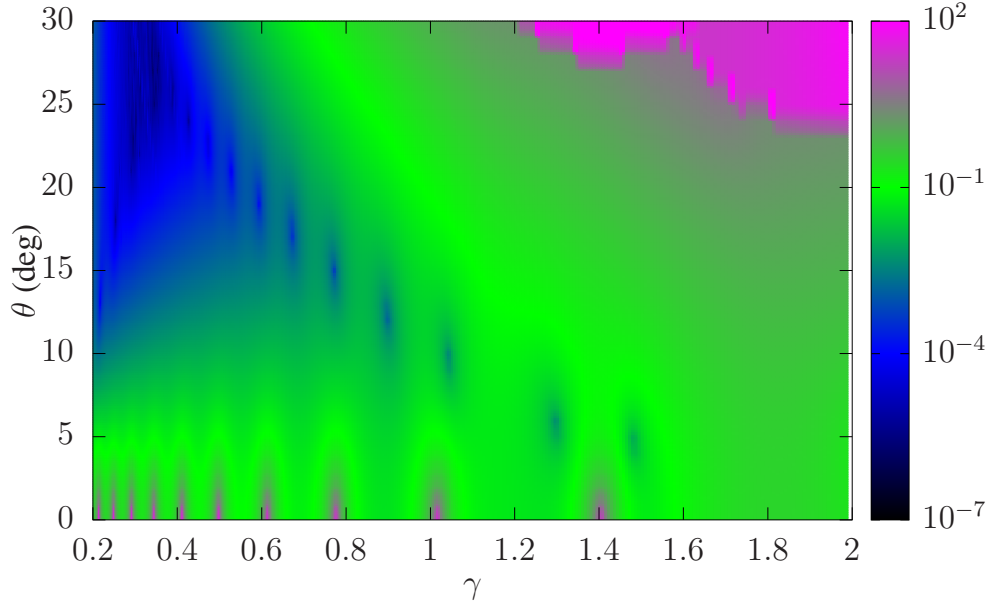}
\caption{
Dependence of the $s$ function (in a.u.) given by Eq.~\eqref{eq:ds} on the $\theta$ and $\gamma$ parameters for the $2s^2$ state of the He-like carbon ($Z = 6$) ion.
The CSFs are constructed from one-electron $s$ and $p$ Dirac orbitals obtained in the basis of $30$ B-splines. 
The size of the spherical box was chosen to be 15 a.u.
}
\label{fig:3d}
\end{figure}

%
Let us start with a brief description of the principles of the stabilization and basis balancing methods, which are applied to the conventional (hermitian) Hamiltonian.
In the SM~\cite{Holoien_Midtal}, the basis-set parameters are chosen in such a way to provide a minimal value for the rate of change of the energy with respect to a variation of these parameters.
In the framework of the BBM~\cite{Yerokhin_PRA96_042505:2017}, one needs to manipulate the basis to place the resonance just in the middle between the closest quasi-continuum states in the energy scale.
Both these methods utilize the advantages of the finite basis set constructed from the square-integrable functions.
As already was mentioned, such basis set functions cannot properly describe the contribution of the continuum to the autoionizing states.
That is expressed in the energy shift of the state from the exact value.
The size of this shift is, however, strongly resonance-dependent and may be negligible in some cases.
Here we estimate the difference between the results of the complex scaling approach with ones from the stabilization and basis balancing methods considering the state which is known to be significantly coupled with the continuum, namely, $2s^2$ autoionizing state of the He-like carbon ion ($Z = 6$).
For this purpose, we choose the radial grid, which uniquely defines the basis functions constructed from the B-splines, as in Ref.~\cite{Yerokhin_PRA96_042505:2017}:
\begin{equation}
t_i = t_0 e^{A(i/N)^\gamma},
\label{eq:grid}
\end{equation}
where $A = \ln{\left(t_{\rm max} / t_0\right)}$, $t_{\rm max}$ is the radial size of the spherical cavity, $t_0$ is the radius of the nucleus, and $\gamma$ is the basis set parameter.
The energies of the autoionizing and quasi-continuum states depend strongly on the parameter~$\gamma$ and form $\gamma$-parametric trajectories, which are analyzed in accordance with the SM and BBM.
For the sake of simplicity, we include only the CSFs being constructed from one-electron $s$ and $p$ Dirac orbitals.
Fig.~\ref{fig:stabilization} presents the $\gamma$-parametric energy trajectories for the $2s^2$ state of the He-like carbon ($Z = 6$) ion obtained in the basis of $30$ B-splines.
This figure also presents the energies obtained with the usage of the SM and BBM for each $\gamma$-parametric trajectory.
From Fig.~\ref{fig:stabilization}, it is seen that at $\gamma$ smaller than $0.5$ the results of the SM and BBM are very close to each other.
Before we proceed to the investigation of the convergence with respect to the number of B-splines, let us explore the dependence of the results obtained within the CS approach on the $\gamma$ parameter.
%
%
\\ \indent
%
%
As was discussed in the preceding section, in the uniform complex rotation approach, the Hamiltonian depends on the $\theta$ parameter.
Energies of the bound and quasi-bound states in this method are, however, $\theta$-independent for $\theta_c \leqslant \theta < \pi/2$ where $\theta_c$ is the critical angle given by~\cite{Balslev_CMP22_280:1971, Simon}
\begin{equation}
\theta_c = \arctan{\left[\Gamma / (2\left(E - E_t\right))\right]}.
\end{equation}
Here $\Gamma$ and $E$ are the Auger width and energy of the level of interest, respectively, and $E_t$ is the autoionization threshold energy, which for the $2s^2$ state is provided by the ground state of the corresponding H-like ion.
It should be noted that the energies do not depend on $\theta$ only if the complete or large basis set is utilized.
In practice, however, one has to deal with an incomplete basis set that requires a search of an optimal angle for the uniform complex rotation.
This angle corresponds to the stationary point of the $\theta$-parametric energy curve in the complex plane.
In our case, one needs to find the stationary point of the $(\gamma,\theta)$-parametric energy surface in the complex plane.
That is equivalent to the search for the minimum of the function
\begin{equation}
s (\gamma, \theta) \equiv \sqrt{\left\vert\frac{dE}{d\theta}\right\vert^2 + \left\vert\frac{dE}{d\gamma}\right\vert^2}.
\label{eq:ds}
\end{equation}
%
Fig.~\ref{fig:3d} presents the $s$ function~\eqref{eq:ds} for the $2s^2$ state of the He-like carbon ($Z = 6$) ion obtained in the basis of $30$ B-splines.
From this figure it is seen that the $s(\gamma, \theta)$ function takes minimal values at $\gamma$ from $0.3$ to $0.5$ and $\theta$ from $20^\circ$ to $30^\circ$.
For $\gamma$ and $\theta$ changing within this area, the energy of the $2s^2$ state exhibits very stable behavior.
%
%
\\ \indent
%
%
We now turn to the investigation of the convergence of the results obtained within the SM, BBM, and CS methods with respect to the number of basis functions.
Table~\ref{tb:sm_bbm_cs} presents the energy of the $2s^2$ state of the He-like C ($Z=6$) ion for different numbers of the B-splines.
\begin{table}[H]
\centering
\caption{
Energy (in a.u.) of the $2s^2$ state of the He-like C ($Z = 6$) ion obtained within the stabilization method (SM), the basis balancing method (BBM), and the complex scaling (CS) approach. 
The CSFs are constructed from one-electron $s$ and $p$ Dirac orbitals obtained in the basis of $N$ functions. 
The size of the spherical box was chosen to be 15 a.u.
Parameter $\gamma$ is varied in the range between $0.3$ and $0.5$.
The calculations within the CS approach are performed for $\theta$ varying from $20^\circ$ to $25^\circ$.
\label{tb:sm_bbm_cs}
}
\begin{tabular}{c
                S[table-format= -1.5(1)] 
                S[table-format= -1.5(1)] 
                S[table-format= -1.8(2)]
                S[table-format= -1.5(2)]}
\hline\hline
$N$ &     {SM}   &     {BBM}   & \multicolumn{2}{c}{CS}  \\
    &            &             & {${\rm Re}\left(E\right)$} & {${\rm Im}\left(E\right) \times 10^3$}  \\
\hline
30 & -8.29130(3) & -8.2924(2)  & -8.291450(4)   & -3.529(4)    \\
40 & -8.29134(4) & -8.2921(1)  & -8.2914499(2)  & -3.5290(3)   \\
50 & -8.29137(3) & -8.29197(3) & -8.29144998(9) & -3.52906(15) \\
60 & -8.29140(2) & -8.2919(1)  & -8.29144998(9) & -3.52907(13) \\
\hline\hline
\end{tabular}
\end{table}
\noindent
The calculations within the stabilization and basis balancing methods are performed for the $\gamma$ parameter 
varying from $0.3$ to $0.5$.
In Table~\ref{tb:sm_bbm_cs}, we present the average values of the energies originating from different energetic curves 
corresponding to this $\gamma$ interval (see Fig.~\ref{fig:stabilization}).
The uncertainty reflects the dependence of the results on the choice of the curve.
From Table~\ref{tb:sm_bbm_cs}, it is seen that the BBM results stronger depend on the energetic curve than the SM ones.
It can be due to the fact that in the BBM the resonance position is balanced with respect to the closest quasi-continuum states whereas in the SM the whole spectra is effectively taken into account.
For both methods, the dependence on the energetic curve
strongly masks the convergence with respect to the number of basis functions and gives the main source of the uncertainty.
The calculations within the CS approach are performed for $\gamma$ varying from $0.3$ to $0.5$ and $\theta$ varying in the range between $20^\circ$ and $25^\circ$.
The dependence of the energy on the $\gamma$ and $\theta$ parameters forms the uncertainty indicated in Table~\ref{tb:sm_bbm_cs}.
It is seen that the energy obtained within the CS approach exhibits extremely fast convergence with respect to the number of basis functions.
It is also seen that the energies obtained within the SM and BBM differ from the one calculated using the complex scaling approach by more than $1$ meV and $10$ meV, respectively, the values which actually define accuracy limits of the SM and BBM.
We note also that, working with SM and BBM, one needs to re-select the basis set parameters each time when
the number of the basis functions is enlarged. 
The necessity of this procedure drastically increase the number of required computation time and, thus, strongly reduces the advantage of the real arithmetic.
%
%
\subsection{Energies and Auger widths of the $LL$ resonances}
%
%
We now apply the configuration-interaction complex-scaling method for the calculation of the energies and Auger width of $LL$ resonances of the He-like ions from boron ($Z = 5$) to argon ($Z = 18$).
The simplicity of the system studied allows performing the full CI calculations, i.e. the configuration space is formed from all possible combinations of the one-electron Dirac orbitals appearing for a given number of the B-splines.
In the present paper, the B-splines of order $11$ are utilized. 
Such a high order of the B-splines is chosen to guarantee the correct behavior of the one-electron Dirac orbitals with orbital angular momenta up to $L = 8$ at the origin.
The one-electron orbitals with proper behavior at the origin appear to be less dependent on the choice of the complex rotation angle $\theta$ and, thus, provide more accurate results.
The accuracy of the DCB eigenvalues apart from the choice of the $\theta$ and $\gamma$ parameters depends on the number of B-splines and the number of the orbital angular momenta $L$ included.
In order to estimate the uncertainty arising from the number of orbital angular momenta we carry out the CI calculations for $L \leqslant 8$ and estimate the tail contributions via polynomial least square fitting of the increments in powers of $1/L$ as in Refs.~\cite{Yerokhin_PRA86_042507:2012, Yerokhin_PRA96_042505:2017, Kaygorodov_PRA99_032505:2019}.
An example of such uncertainty analysis is presented in Table~\ref{tab:mult} for the $2s^2$ state of the carbon ($Z = 6$) ion.
\begin{table}[H]
\centering
\caption{
Energy~$E$ and Auger width~$\Gamma_{\rm Aug}$ of the $2s^2$ state of the He-like carbon ($Z = 6$) ion obtained within the configuration-interaction complex-scaling method. 
The CSFs are constructed from one-electron Dirac orbitals with orbital angular momenta up to $L_{\rm max}$ being obtained in the basis of $N$ B-splines. 
The size of the spherical box was chosen to be 15 a.u., $\gamma = 0.3$, and $\theta = 20^\circ$.
The values listed after the second row are the increments obtained on successively adding configurations while increasing $L_{\rm max}$.
\label{tab:mult}}
\begin{tabular}{l|
                S[table-format= -1.7(1)] 
                S[table-format= -1.7(1)] 
                S[table-format= -1.7(1)]|
                S[table-format= -1.5(1)] 
                S[table-format= -1.5(1)] 
                S[table-format= -1.5(1)]}
\hline\hline
$L_{\rm max}$ & \multicolumn{3}{c}{$E$ [a.u.]}       & \multicolumn{3}{c}{$\Gamma_{\rm Aug} \times 10^3$ [a.u.]}              \\
              & {$N = 30$} & {$N = 40$} & {$N = 50$} & {$N = 30$} & {$N = 40$} & {$N = 50$} \\
\hline
           1 & -8.2914506    & -8.2914500    & -8.2914499    &  7.05659    &  7.05801    &  7.05800 \\
           2 & -0.0007961    & -0.0007960    & -0.0007960    & -0.08175    & -0.08155    & -0.08152 \\
           3 & -0.0001180    & -0.0001183    & -0.0001183    & -0.01888    & -0.01872    & -0.01870 \\
           4 & -0.0000375    & -0.0000378    & -0.0000378    & -0.00672    & -0.00658    & -0.00656 \\
           5 & -0.0000159    & -0.0000161    & -0.0000162    & -0.00302    & -0.00291    & -0.00289 \\
           6 & -0.0000078    & -0.0000081    & -0.0000081    & -0.00157    & -0.00149    & -0.00148 \\
           7 & -0.0000043    & -0.0000045    & -0.0000045    & -0.00091    & -0.00085    & -0.00083 \\
           8 & -0.0000025    & -0.0000027    & -0.0000027    & -0.00057    & -0.00052    & -0.00051 \\
9-$\infty$   & -0.0000057 & -0.0000068 & -0.0000071 & -0.00176 & -0.00149 & -0.00142 \\
Total        & -8.2924383 & -8.2924402 & -8.2924407 &  6.94141 &  6.94391 &  6.94407 \\
\hline\hline
\end{tabular}
\end{table}
\noindent
From this table, it is seen that for the basis of more than $40$ B-splines the dominant contribution to the uncertainty of the DCB eigenvalues is provided by the configuration states with orbital angular momenta $L \geqslant 9$, whose contributions are taken into account by extrapolation.
Therefore, in what follows we solve the complex rotated DCB equation in the configuration space formed from all possible combinations of the one-electron Dirac orbitals constructed out of $40$ or $50$ B-splines.
%
%
\\ \indent
%
%
In order to obtain the energies of the $LL$ resonances with an accuracy at a few meV level, we supplement the solutions of the complex rotated DCB equation with the nuclear recoil and QED corrections.
Both corrections are obtained with the usage of the conventional (hermitian) DCB Hamiltonian.
%
%
The nuclear recoil effect arising due to the finite nuclear mass $M$ admits fully relativistic treatment only within the framework of QED~\cite{Shabaev_recoil, Adkins_PRA76_042508:2007}.
Here we account for this effect in the lowest-order relativistic approximation and to first order in $m/M$ via the inclusion of the mass shift operator~\cite{Shabaev_recoil, Palmer_JPB20_5987:1987}
\begin{equation}
H_{\rm MS} = \frac{1}{2M}\sum_{i,j}\left\lbrace
\bp_i\cdot\bp_j - \frac{\alpha Z}{r_i}\left[\balpha_i + \frac{\left(\balpha_i\cdot\br_i\right)\br_i}{r_i^2}\right]\cdot\bp_j
\right\rbrace,
\end{equation}
into the DCB Hamiltonian.
The nuclear recoil correction to the energy of the particular $LL$ resonance is given by the first-order perturbation theory with respect to this additional term~\cite{Tupitsyn_PRA68_022511:2003}.
%
As already mentioned, in addition to the nuclear recoil corrections we supplement the complex rotated DCB energies with the QED corrections.
The {\it ab initio} evaluation of these corrections still remains a challenging task even for He-like systems for which the methods of the QED calculations are currently well established (see, e.g.,~\cite{Artemyev_PRA71_062104:2005, 
 Pachucki_PRA81_022507:2010, Malyshev_PRA99_010501:2019} and references therein).
It is also worth to mention that to the best of our knowledge no attempt was made to compute the two-electron QED effects on the energies of the autoionizing states.
In the present paper, we evaluate the QED corrections utilizing the model QED operator~\cite{Shabaev_PRA88_012513:2013}, constructed with the usage of the \texttt{QEDMOD} package~\cite{Shabaev_QEDMOD}.  
We evaluate the QED correction as the difference between the CI results obtained with and without the model QED operator included into the DCB Hamiltonian.
This approach has shown its efficiency in numerous investigations~\cite{Tupitsyn_PRL117_253001:2016, Yerokhin_PRA96_042505:2017, Yerokhin_JPCRD47_023105:2018, Kaygorodov_PRA99_032505:2019}.
However, in the QED model operator method, the screened QED corrections are taken into account only approximately.
These corrections as well as the QED part of the two-photon-exchange contributions give rise to another source of uncertainty.
We also note that the frequency-dependent Breit correction was found to be of minor importance for systems under investigation and, therefore, its contribution can be omitted.
%
%
\\ \indent
%
%
Table~\ref{tab:long} presents the energies and Auger widths of the $LL$ resonances of the He-like ions from boron ($Z = 5$) to argon ($Z = 18$).
In this table, the complex rotated DCB energy, the QED correction, and the nuclear recoil correction are explicitly shown. 
The presented Auger widths $\Gamma_{\rm Aug}$ were calculated only by means of the CS DCB Hamiltonian. 
The smallness of the Auger widths of the $2p_{1/2}^2\,(J=0)$, $2p_{3/2}^2\,(J=2)$, and $2p_{1/2}2p_{3/2}\,(J=1)$ resonances
is explained by the fact that the Auger decay of the ${}^3P_0$, ${}^3P_2$, and ${}^3P_1$ states corresponding to these resonances in the $LS$-coupling scheme, respectively, is strictly forbidden in the nonrelativistic limit.
Energies $E_{\rm tot}$ are supplemented with the total uncertainties from all calculated contributions as well as from uncalculated high-order QED corrections.
The uncertainty due to the uncalculated QED corrections was estimated by analysis of the related contributions for the ground and single-excited states in He-like ions~\cite{Artemyev_PRA71_062104:2005}.
In most cases, the accuracy of the present calculations is limited by the uncertainties from the QED contributions.
Using the presented results with the available high-precision data for the energies of the ground and lowest excited states 
(see Refs.~\cite{Artemyev_PRA71_062104:2005, Pachucki_PRA81_022507:2010,
Malyshev_PRA99_010501:2019}), one can easily find the corresponding transition
energies.
%
\setlength\LTcapwidth{\linewidth}
\begingroup
\begin{center}
\begin{longtable}{ccc|
 S[table-format= -3.7]
 S[table-format=  2.7]
 S[table-format= -2.7]
 S[table-format= -3.7(2)]
 l}
\caption{
Energies $E_{\rm tot}$ and Auger widths~$\Gamma_{\rm Aug}$ of the $LL$ resonances of the He-like ions from boron ($Z = 5$) to argon ($Z = 18$), in a.u. 
The CS DCB energy, the QED correction, and the nuclear recoil correction are explicitly shown. 
Energies $E_{\rm tot}$ are supplemented with the total uncertainties from all calculated and 
uncalculated  contributions. The nuclear charge radii are taken from Ref.~\cite{Angeli}.
\label{tab:long}
}
\\
\hline
\hline
Ion & Resonance  & $J$ & \multicolumn{1}{c}{DCB} & \multicolumn{1}{c}{Recoil} & \multicolumn{1}{c}{QED} & \multicolumn{1}{c}{$E_{\rm tot}$} & \multicolumn{1}{c}{$\Gamma_{\rm Aug}$} \\
\endfirsthead
\caption*{Table \ref{tab:long} ({\it Continued.})}\\\toprule
\hline
 Ion & {Resonance}  & $J$ & \multicolumn{1}{c}{DCB} & \multicolumn{1}{c}{Recoil} & {QED} & {$E_{\rm tot}$} & {$\Gamma_{\rm Aug}$} \\
\hline
\endhead
\hline
$^{11}$B$^{3+}$ &       $2s_{1/2}^2$ & 0 &     -5.6628771 &                 0.0002821 &                 0.0000856 &   -5.662509(24) & 6.674(1)$\times 10^{-3}$ \\ 

  &       $2p_{1/2}^2$ & 0 &     -5.4702350 &                 0.0002805 &                -0.0000007 &  -5.4699552(88) &     $< 10^{-6}$ \\ 

  &       $2p_{3/2}^2$ & 0 &     -5.1457619 &                 0.0002760 &                 0.0000203 &   -5.145466(42) & 3.0(2)$\times 10^{-4}$ \\ 

  &                    & 2 &     -5.4694360 &                 0.0002804 &                 0.0000013 &  -5.4691543(48) & $< 2 \times10^{-6}$ \\ 

  & $2s_{1/2}2p_{1/2}$ & 0 &     -5.6151794 &                 0.0002811 &                 0.0000541 &   -5.614844(13) & 3.314(7)$\times 10^{-4}$ \\ 

  &                    & 1 &     -5.6149173 &                 0.0002811 &                 0.0000548 &   -5.614582(13) & 3.277(7)$\times 10^{-4}$ \\ 

  & $2s_{1/2}2p_{3/2}$ & 1 &     -5.3817344 &                 0.0002710 &                 0.0000465 &   -5.381417(54) & 3.09(4)$\times 10^{-3}$ \\ 

  &                    & 2 &     -5.6143278 &                 0.0002810 &                 0.0000562 &   -5.613991(13) & 3.241(5)$\times 10^{-4}$ \\ 

  & $2p_{1/2}2p_{3/2}$ & 1 &     -5.4699408 &                 0.0002805 &                       0.0 &  -5.4696604(36) &     $< 10^{-6}$ \\ 

  &                    & 2 &     -5.4044733 &                 0.0002711 &                 0.0000018 &   -5.404200(30) & 5.52(1)$\times 10^{-3}$ \\ 

$^{12}$C$^{4+}$ &       $2s_{1/2}^2$ & 0 &     -8.2924407 &                 0.0003786 &                 0.0001679 &   -8.291894(41) & 6.944(1)$\times 10^{-3}$ \\ 

  &       $2p_{1/2}^2$ & 0 &     -8.0579693 &                 0.0003770 &                -0.0000015 &   -8.057594(13) & $< 2 \times10^{-7}$ \\ 

  &       $2p_{3/2}^2$ & 0 &     -7.6535338 &                 0.0003724 &                 0.0000413 &   -7.653120(54) & 3.1(1)$\times 10^{-4}$ \\ 

  &                    & 2 &     -8.0562426 &                 0.0003770 &                 0.0000028 &  -8.0558628(66) & 6.9(9)$\times 10^{-7}$ \\ 

  & $2s_{1/2}2p_{1/2}$ & 0 &     -8.2349686 &                 0.0003777 &                 0.0001058 &   -8.234485(22) & 3.377(6)$\times 10^{-4}$ \\ 

  &                    & 1 &     -8.2343998 &                 0.0003776 &                 0.0001072 &   -8.233915(22) & 3.32(1)$\times 10^{-4}$ \\ 

  & $2s_{1/2}2p_{3/2}$ & 1 &     -7.9435675 &                 0.0003661 &                 0.0000940 &   -7.943107(71) & 3.35(4)$\times 10^{-3}$ \\ 

  &                    & 2 &     -8.2331403 &                 0.0003776 &                 0.0001102 &   -8.232653(22) & 3.27(1)$\times 10^{-4}$ \\ 

  & $2p_{1/2}2p_{3/2}$ & 1 &     -8.0573405 &                 0.0003770 &                       0.0 &  -8.0569635(38) &     $< 10^{-6}$ \\ 

  &                    & 2 &     -7.9712525 &                 0.0003667 &                 0.0000031 &   -7.970883(40) & 6.017(9)$\times 10^{-3}$ \\ 

$^{14}$N$^{5+}$ &       $2s_{1/2}^2$ & 0 &    -11.4234480 &                 0.0004470 &                 0.0002953 &  -11.422706(64) & 7.146(1)$\times 10^{-3}$ \\ 

  &       $2p_{1/2}^2$ & 0 &    -11.1468905 &                 0.0004456 &                -0.0000027 &  -11.146448(19) & 1.1(6)$\times 10^{-7}$ \\ 

  &       $2p_{3/2}^2$ & 0 &    -10.6620276 &                 0.0004412 &                 0.0000745 &  -10.661512(64) & 3.3(2)$\times 10^{-4}$ \\ 

  &                    & 2 &    -11.1435964 &                 0.0004455 &                 0.0000054 & -11.1431455(93) & 1.8(6)$\times 10^{-6}$ \\ 

  & $2s_{1/2}2p_{1/2}$ & 0 &    -11.3564130 &                 0.0004462 &                 0.0001856 &  -11.355781(34) & 3.444(2)$\times 10^{-4}$ \\ 

  &                    & 1 &    -11.3553256 &                 0.0004462 &                 0.0001883 &  -11.354691(34) & 3.363(6)$\times 10^{-4}$ \\ 

  & $2s_{1/2}2p_{3/2}$ & 1 &    -11.0062113 &                 0.0004343 &                 0.0001689 &  -11.005608(88) & 3.54(4)$\times 10^{-3}$ \\ 

  &                    & 2 &    -11.3529409 &                 0.0004461 &                 0.0001939 &  -11.352301(34) & 3.289(5)$\times 10^{-4}$ \\ 

  & $2p_{1/2}2p_{3/2}$ & 1 &    -11.1456981 &                 0.0004456 &                       0.0 & -11.1452525(40) &     $< 10^{-6}$ \\ 

  &                    & 2 &    -11.0385634 &                 0.0004352 &                 0.0000051 &  -11.038123(40) & 6.39(2)$\times 10^{-3}$ \\ 

$^{16}$O$^{6+}$ &       $2s_{1/2}^2$ & 0 &    -15.0564866 &                 0.0005155 &                 0.0004802 &  -15.055491(96) & 7.304(1)$\times 10^{-3}$ \\ 

  &       $2p_{1/2}^2$ & 0 &    -14.7374643 &                 0.0005142 &                -0.0000045 &  -14.736955(27) & 2.3(7)$\times 10^{-7}$ \\ 

  &       $2p_{3/2}^2$ & 0 &    -14.1716603 &                 0.0005099 &                 0.0001233 &  -14.171027(76) & 3.4(2)$\times 10^{-4}$ \\ 

  &                    & 2 &    -14.7317238 &                 0.0005140 &                 0.0000094 &  -14.731200(14) & 3.9(3)$\times 10^{-6}$ \\ 

  & $2s_{1/2}2p_{1/2}$ & 0 &    -14.9801556 &                 0.0005148 &                 0.0003009 &  -14.979340(50) & 3.516(5)$\times 10^{-4}$ \\ 

  &                    & 1 &    -14.9782594 &                 0.0005148 &                 0.0003056 &  -14.977439(50) & 3.409(6)$\times 10^{-4}$ \\ 

  & $2s_{1/2}2p_{3/2}$ & 1 &    -14.5702588 &                 0.0005027 &                 0.0002790 &   -14.56948(11) & 3.69(5)$\times 10^{-3}$ \\ 

  &                    & 2 &    -14.9741209 &                 0.0005146 &                 0.0003152 &  -14.973291(50) & 3.304(5)$\times 10^{-4}$ \\ 

  & $2p_{1/2}2p_{3/2}$ & 1 &    -14.7353908 &                 0.0005141 &                 0.0000003 & -14.7348764(44) &     $< 10^{-6}$ \\ 

  &                    & 2 &    -14.6067695 &                 0.0005037 &                 0.0000080 &  -14.606258(44) & 6.67(3)$\times 10^{-3}$ \\ 

$^{19}$F$^{7+}$ &       $2s_{1/2}^2$ & 0 &    -19.1922303 &                 0.0005533 &                 0.0007355 &   -19.19094(14) & 7.434(3)$\times 10^{-3}$ \\ 

  &       $2p_{1/2}^2$ & 0 &    -18.8302321 &                 0.0005522 &                -0.0000069 &  -18.829687(37) & 4.8(6)$\times 10^{-7}$ \\ 

  &       $2p_{3/2}^2$ & 0 &    -18.1828693 &                 0.0005482 &                 0.0001910 &  -18.182130(88) & 3.5(3)$\times 10^{-4}$ \\ 

  &                    & 2 &    -18.8208917 &                 0.0005519 &                 0.0000153 &  -18.820324(19) & 8(1)$\times 10^{-6}$ \\ 

  & $2s_{1/2}2p_{1/2}$ & 0 &    -19.1069359 &                 0.0005528 &                 0.0004597 &  -19.105923(70) & 3.600(7)$\times 10^{-4}$ \\ 

  &                    & 1 &    -19.1038519 &                 0.0005527 &                 0.0004672 &  -19.102832(70) & 3.463(7)$\times 10^{-4}$ \\ 

  & $2s_{1/2}2p_{3/2}$ & 1 &    -18.6363161 &                 0.0005411 &                 0.0004325 &   -18.63534(13) & 3.81(4)$\times 10^{-3}$ \\ 

  &                    & 2 &    -19.0971277 &                 0.0005525 &                 0.0004828 &  -19.096092(70) & 3.326(8)$\times 10^{-4}$ \\ 

  & $2p_{1/2}2p_{3/2}$ & 1 &    -18.8268543 &                 0.0005521 &                 0.0000007 & -18.8263015(50) & $< 2 \times10^{-6}$ \\ 

  &                    & 2 &    -18.6762343 &                 0.0005422 &                 0.0000123 &  -18.675680(50) & 6.90(4)$\times 10^{-3}$ \\ 

$^{20}$Ne$^{8+}$ &       $2s_{1/2}^2$ & 0 &    -23.8314470 &                 0.0006527 &                 0.0010752 &   -23.82972(19) & 7.542(1)$\times 10^{-3}$ \\ 

  &       $2p_{1/2}^2$ & 0 &    -23.4258147 &                 0.0006516 &                -0.0000100 &  -23.425173(50) & 9.5(4)$\times 10^{-7}$ \\ 

  &       $2p_{3/2}^2$ & 0 &    -22.6961226 &                 0.0006474 &                 0.0002811 &   -22.69519(10) & 3.6(2)$\times 10^{-4}$ \\ 

  &                    & 2 &    -23.4114149 &                 0.0006512 &                 0.0000235 &  -23.410740(28) & 1.47(5)$\times 10^{-5}$ \\ 

  & $2s_{1/2}2p_{1/2}$ & 0 &    -23.7375939 &                 0.0006522 &                 0.0006701 &  -23.736272(96) & 3.690(4)$\times 10^{-4}$ \\ 

  &                    & 1 &    -23.7328442 &                 0.0006521 &                 0.0006815 &  -23.731511(96) & 3.518(3)$\times 10^{-4}$ \\ 

  & $2s_{1/2}2p_{3/2}$ & 1 &    -23.2050294 &                 0.0006397 &                 0.0006381 &   -23.20375(16) & 3.91(4)$\times 10^{-3}$ \\ 

  &                    & 2 &    -23.7224673 &                 0.0006518 &                 0.0007056 &  -23.721110(96) & 3.346(7)$\times 10^{-4}$ \\ 

  & $2p_{1/2}2p_{3/2}$ & 1 &    -23.4205846 &                 0.0006514 &                 0.0000016 & -23.4199316(61) & $< 2 \times10^{-6}$ \\ 

  &                    & 2 &    -23.2473363 &                 0.0006409 &                 0.0000186 &  -23.246677(59) & 7.08(4)$\times 10^{-3}$ \\ 

$^{13}$Na$^{9+}$ &       $2s_{1/2}^2$ & 0 &     -28.975002 &                  0.000690 &                  0.001514 &   -28.97280(25) & 7.635(1)$\times 10^{-3}$ \\ 

  &       $2p_{1/2}^2$ & 0 &     -28.524916 &                  0.000689 &                 -0.000013 &  -28.524240(66) & 1.72(8)$\times 10^{-6}$ \\ 

  &       $2p_{3/2}^2$ & 0 &     -27.711922 &                  0.000685 &                  0.000397 &   -27.71084(12) & 3.7(2)$\times 10^{-4}$ \\ 

  &                    & 2 &     -28.503662 &                  0.000688 &                  0.000035 &  -28.502939(39) & 2.7(1)$\times 10^{-5}$ \\ 

  & $2s_{1/2}2p_{1/2}$ & 0 &     -28.873071 &                  0.000690 &                  0.000941 &   -28.87144(13) & 3.790(5)$\times 10^{-4}$ \\ 

  &                    & 1 &     -28.866072 &                  0.000690 &                  0.000957 &   -28.86443(13) & 3.581(2)$\times 10^{-4}$ \\ 

  & $2s_{1/2}2p_{3/2}$ & 1 &     -28.277097 &                  0.000677 &                  0.000905 &   -28.27551(20) & 3.99(4)$\times 10^{-3}$ \\ 

  &                    & 2 &     -28.850706 &                  0.000689 &                  0.000993 &   -28.84902(13) & 3.367(3)$\times 10^{-4}$ \\ 

  & $2p_{1/2}2p_{3/2}$ & 1 &     -28.517138 &                  0.000689 &                  0.000003 & -28.5164466(78) & $< 2 \times10^{-6}$ \\ 

  &                    & 2 &     -28.320470 &                  0.000679 &                  0.000028 &  -28.319763(70) & 7.23(4)$\times 10^{-3}$ \\ 

$^{24}$Mg$^{10+}$ &       $2s_{1/2}^2$ & 0 &     -34.623863 &                  0.000790 &                  0.002068 &   -34.62101(33) & 7.716(1)$\times 10^{-3}$ \\ 

  &       $2p_{1/2}^2$ & 0 &     -34.128326 &                  0.000789 &                 -0.000017 &  -34.127554(85) & 2.91(7)$\times 10^{-6}$ \\ 

  &       $2p_{3/2}^2$ & 0 &     -33.230798 &                  0.000785 &                  0.000542 &   -33.22947(14) & 3.9(2)$\times 10^{-4}$ \\ 

  &                    & 2 &     -34.098066 &                  0.000788 &                  0.000049 &  -34.097229(54) & 4.49(6)$\times 10^{-5}$ \\ 

  & $2s_{1/2}2p_{1/2}$ & 0 &     -34.514412 &                  0.000790 &                  0.001281 &   -34.51234(17) & 3.901(4)$\times 10^{-4}$ \\ 

  &                    & 1 &     -34.504470 &                  0.000790 &                  0.001304 &   -34.50238(17) & 3.652(5)$\times 10^{-4}$ \\ 

  & $2s_{1/2}2p_{3/2}$ & 1 &     -33.853273 &                  0.000777 &                  0.001243 &   -33.85125(24) & 4.06(4)$\times 10^{-3}$ \\ 

  &                    & 2 &     -34.482472 &                  0.000789 &                  0.001355 &   -34.48033(17) & 3.387(7)$\times 10^{-4}$ \\ 

  & $2p_{1/2}2p_{3/2}$ & 1 &     -34.117134 &                  0.000789 &                  0.000005 &  -34.116340(10) & $< 2 \times10^{-6}$ \\ 

  &                    & 2 &     -33.896040 &                  0.000778 &                  0.000040 &  -33.895222(86) & 7.34(4)$\times 10^{-3}$ \\ 

$^{27}$Al$^{11+}$ &       $2s_{1/2}^2$ & 0 &     -40.779109 &                  0.000827 &                  0.002752 &   -40.77553(42) & 7.788(1)$\times 10^{-3}$ \\ 

  &       $2p_{1/2}^2$ & 0 &     -40.236929 &                  0.000826 &                 -0.000020 &   -40.23612(11) & 4.66(6)$\times 10^{-6}$ \\ 

  &       $2p_{3/2}^2$ & 0 &     -39.253305 &                  0.000822 &                  0.000718 &   -39.25177(17) & 4.0(2)$\times 10^{-4}$ \\ 

  &                    & 2 &     -40.195135 &                  0.000825 &                  0.000067 &  -40.194243(72) & 7.4(1)$\times 10^{-5}$ \\ 

  & $2s_{1/2}2p_{1/2}$ & 0 &     -40.662765 &                  0.000827 &                  0.001699 &   -40.66024(21) & 4.025(8)$\times 10^{-4}$ \\ 

  &                    & 1 &     -40.649071 &                  0.000827 &                  0.001730 &   -40.64651(21) & 3.738(6)$\times 10^{-4}$ \\ 

  & $2s_{1/2}2p_{3/2}$ & 1 &     -39.934371 &                  0.000814 &                  0.001661 &   -39.93190(29) & 4.12(5)$\times 10^{-3}$ \\ 

  &                    & 2 &     -40.618454 &                  0.000826 &                  0.001801 &   -40.61583(21) & 3.412(7)$\times 10^{-4}$ \\ 

  & $2p_{1/2}2p_{3/2}$ & 1 &     -40.221253 &                  0.000826 &                  0.000009 &  -40.220418(14) & $< 2 \times10^{-6}$ \\ 

  &                    & 2 &     -39.974458 &                  0.000816 &                  0.000058 &   -39.97358(10) & 7.42(4)$\times 10^{-3}$ \\ 

$^{28}$Si$^{12+}$ &       $2s_{1/2}^2$ & 0 &     -47.441930 &                  0.000928 &                  0.003583 &   -47.43742(52) & 7.853(1)$\times 10^{-3}$ \\ 

  &       $2p_{1/2}^2$ & 0 &     -46.851709 &                  0.000927 &                 -0.000021 &   -46.85080(13) & 7.1(1)$\times 10^{-6}$ \\ 

  &       $2p_{3/2}^2$ & 0 &     -45.780017 &                  0.000923 &                  0.000928 &   -45.77817(19) & 4.1(2)$\times 10^{-4}$ \\ 

  &                    & 2 &     -46.795465 &                  0.000926 &                  0.000089 &  -46.794451(95) & 1.16(1)$\times 10^{-4}$ \\ 

  & $2s_{1/2}2p_{1/2}$ & 0 &     -47.319385 &                  0.000928 &                  0.002206 &   -47.31625(26) & 4.157(2)$\times 10^{-4}$ \\ 

  &                    & 1 &     -47.301016 &                  0.000928 &                  0.002247 &   -47.29784(26) & 3.84(1)$\times 10^{-4}$ \\ 

  & $2s_{1/2}2p_{3/2}$ & 1 &     -46.521263 &                  0.000914 &                  0.002171 &   -46.51818(34) & 4.17(5)$\times 10^{-3}$ \\ 

  &                    & 2 &     -47.259405 &                  0.000927 &                  0.002343 &   -47.25614(26) & 3.437(9)$\times 10^{-4}$ \\ 

  & $2p_{1/2}2p_{3/2}$ & 1 &     -46.830238 &                  0.000926 &                  0.000014 &  -46.829298(20) & 6(3)$\times 10^{-7}$ \\ 

  &                    & 2 &     -46.556124 &                  0.000916 &                  0.000082 &   -46.55513(13) & 7.48(4)$\times 10^{-3}$ \\ 

$^{31}$P$^{13+}$ &       $2s_{1/2}^2$ & 0 &     -54.613632 &                  0.000965 &                  0.004579 &   -54.60809(64) & 7.911(1)$\times 10^{-3}$ \\ 

  &       $2p_{1/2}^2$ & 0 &     -53.973757 &                  0.000964 &                 -0.000017 &   -53.97281(16) & 1.05(2)$\times 10^{-5}$ \\ 

  &       $2p_{3/2}^2$ & 0 &     -52.811508 &                  0.000959 &                  0.001175 &   -52.80937(22) & 4.3(2)$\times 10^{-4}$ \\ 

  &                    & 2 &     -53.899759 &                  0.000962 &                  0.000115 &   -53.89868(12) & 1.77(2)$\times 10^{-4}$ \\ 

  & $2s_{1/2}2p_{1/2}$ & 0 &     -54.485633 &                  0.000965 &                  0.002810 &   -54.48186(32) & 4.305(1)$\times 10^{-4}$ \\ 

  &                    & 1 &     -54.461556 &                  0.000964 &                  0.002862 &   -54.45773(32) & 3.946(7)$\times 10^{-4}$ \\ 

  & $2s_{1/2}2p_{3/2}$ & 1 &     -53.614881 &                  0.000952 &                  0.002783 &   -53.61115(40) & 4.21(5)$\times 10^{-3}$ \\ 

  &                    & 2 &     -54.406141 &                  0.000963 &                  0.002991 &   -54.40219(32) & 3.47(2)$\times 10^{-4}$ \\ 

  & $2p_{1/2}2p_{3/2}$ & 1 &     -53.944897 &                  0.000963 &                  0.000021 &  -53.943913(28) & 8(4)$\times 10^{-7}$ \\ 

  &                    & 2 &     -53.641418 &                  0.000953 &                  0.000114 &   -53.64035(17) & 7.50(5)$\times 10^{-3}$ \\ 

$^{32}$S$^{14+}$ &       $2s_{1/2}^2$ & 0 &     -62.295647 &                  0.001066 &                  0.005757 &   -62.28882(78) & 7.963(1)$\times 10^{-3}$ \\ 

  &       $2p_{1/2}^2$ & 0 &     -61.604283 &                  0.001065 &                 -0.000006 &   -61.60322(20) & 1.48(2)$\times 10^{-5}$ \\ 

  &       $2p_{3/2}^2$ & 0 &     -60.348353 &                  0.001060 &                  0.001458 &   -60.34583(26) & 4.5(2)$\times 10^{-4}$ \\ 

  &                    & 2 &     -61.230680 &                  0.001063 &                  0.000144 &   -61.22947(21) & 7.48(5)$\times 10^{-3}$ \\ 

  & $2s_{1/2}2p_{1/2}$ & 0 &     -62.162977 &                  0.001066 &                  0.003522 &   -62.15839(39) & 4.468(1)$\times 10^{-4}$ \\ 

  &                    & 1 &     -62.132053 &                  0.001066 &                  0.003588 &   -62.12740(39) & 4.078(6)$\times 10^{-4}$ \\ 

  & $2s_{1/2}2p_{3/2}$ & 1 &     -61.216210 &                  0.001052 &                  0.003508 &   -61.21165(48) & 4.24(5)$\times 10^{-3}$ \\ 

  &                    & 2 &     -62.059542 &                  0.001064 &                  0.003756 &   -62.05472(39) & 3.50(2)$\times 10^{-4}$ \\ 

  & $2p_{1/2}2p_{3/2}$ & 1 &     -61.566101 &                  0.001064 &                  0.000031 &  -61.565007(38) & 1.0(3)$\times 10^{-6}$ \\ 

  &                    & 2 &     -61.508840 &                  0.001053 &                  0.000157 &   -61.50763(17) & 2.63(3)$\times 10^{-4}$ \\ 

$^{35}$Cl$^{15+}$ &       $2s_{1/2}^2$ & 0 &     -70.489528 &                  0.001103 &                  0.007135 &   -70.48129(94) & 8.012(1)$\times 10^{-3}$ \\ 

  &       $2p_{1/2}^2$ & 0 &     -69.744626 &                  0.001101 &                  0.000017 &   -69.74351(24) & 2.01(2)$\times 10^{-5}$ \\ 

  &       $2p_{3/2}^2$ & 0 &     -68.391105 &                  0.001097 &                  0.001778 &   -68.38823(30) & 4.7(2)$\times 10^{-4}$ \\ 

  &                    & 2 &     -69.324198 &                  0.001099 &                  0.000178 &   -69.32292(25) & 7.43(4)$\times 10^{-3}$ \\ 

  & $2s_{1/2}2p_{1/2}$ & 0 &     -70.352997 &                  0.001103 &                  0.004351 &   -70.34754(47) & 4.641(1)$\times 10^{-4}$ \\ 

  &                    & 1 &     -70.313991 &                  0.001102 &                  0.004433 &   -70.30846(47) & 4.225(8)$\times 10^{-4}$ \\ 

  & $2s_{1/2}2p_{3/2}$ & 1 &     -69.326288 &                  0.001089 &                  0.004357 &   -69.32084(56) & 4.27(6)$\times 10^{-3}$ \\ 

  &                    & 2 &     -70.220551 &                  0.001101 &                  0.004650 &   -70.21480(47) & 3.53(2)$\times 10^{-4}$ \\ 

  & $2p_{1/2}2p_{3/2}$ & 1 &     -69.694787 &                  0.001101 &                  0.000044 &  -69.693642(52) & 1.3(8)$\times 10^{-6}$ \\ 

  &                    & 2 &     -69.623669 &                  0.001090 &                  0.000214 &   -69.62236(23) & 3.79(3)$\times 10^{-4}$ \\ 

$^{40}$Ar$^{16+}$ &       $2s_{1/2}^2$ & 0 &     -79.196961 &                  0.001084 &                  0.008730 &    -79.1871(11) & 8.056(1)$\times 10^{-3}$ \\ 

  &       $2p_{1/2}^2$ & 0 &     -78.396270 &                  0.001083 &                  0.000060 &   -78.39513(28) & 2.66(3)$\times 10^{-5}$ \\ 

  &       $2p_{3/2}^2$ & 0 &     -76.940290 &                  0.001078 &                  0.002134 &   -76.93708(34) & 4.9(3)$\times 10^{-4}$ \\ 

  &                    & 2 &     -77.922194 &                  0.001080 &                  0.000215 &   -77.92090(29) & 7.33(4)$\times 10^{-3}$ \\ 

  & $2s_{1/2}2p_{1/2}$ & 0 &     -79.057381 &                  0.001084 &                  0.005310 &   -79.05099(56) & 4.833(1)$\times 10^{-4}$ \\ 

  &                    & 1 &     -79.008977 &                  0.001084 &                  0.005408 &   -79.00249(56) & 4.400(6)$\times 10^{-4}$ \\ 

  & $2s_{1/2}2p_{3/2}$ & 1 &     -77.946209 &                  0.001071 &                  0.005344 &   -77.93979(64) & 4.30(6)$\times 10^{-3}$ \\ 

  &                    & 2 &     -78.890181 &                  0.001082 &                  0.005685 &   -78.88341(56) & 3.57(2)$\times 10^{-4}$ \\ 

  & $2p_{1/2}2p_{3/2}$ & 1 &     -78.331956 &                  0.001082 &                  0.000061 &  -78.330813(70) & 1.7(7)$\times 10^{-6}$ \\ 

  &                    & 2 &     -78.245356 &                  0.001072 &                  0.000288 &   -78.24400(31) & 5.31(4)$\times 10^{-4}$ \\

\hline\hline
\end{longtable}
\end{center}
\endgroup

In Table~\ref{tab:comparison}, we compare 
some of our results with other
nonrelativistic~\cite{Ho_PRA23_2137:1981} and relativistic 
calculations~\cite{Muller_PRA98_033416:2018}. In Ref.~\cite{Ho_PRA23_2137:1981}, 
the calculations were performed
using the complex scaling technique in combination with Hylleraas-type functions without taking into 
account the QED corrections. 
Since the nonrelativistic method cannot resolve the fine 
structure of the $2s2p$ resonance, for our three values for 
the $2s_{1/2}2p_{1/2} \,(J = 0,1)$ and $2s_{1/2}2p_{3/2} \, (J = 2)$ 
states there is only one corresponding
value of Ref.~\cite{Ho_PRA23_2137:1981}. As one can see 
from the table, our results are in reasonable agreement with the nonrelativistic ones. 
We also compared the values obtained 
for the carbon ion ($Z = 6$) with the recent relativistic 
calculations of Ref.~\cite{Muller_PRA98_033416:2018}. These 
calculations were performed employing the many-body perturbation
theory in an all-order formulation 
with the complex scaling technique (see Ref.~\cite{Lindroth_AQC63_247:2012} and references therein). 
The QED corrections were taken into account using 
the Welton method which is different from 
the QED model operator approach. 
However, the results of Ref.~\cite{Muller_PRA98_033416:2018} are in excellent
agreement with our values. 
\begingroup
\begin{center}
\begin{ThreePartTable}
\begin{longtable}{ccc|
S[table-format= -3.0(2)]
S|
S
S}\\
\caption{The comparison of the calculated energies~$E$ and 
Auger widths~$\Gamma_{\rm Aug}$ of the $LL$ resonances of the He-like ions with 
other nonrelativistic~\cite{Ho_PRA23_2137:1981} and relativistic 
results~\cite{Muller_PRA98_033416:2018}.}
\label{tab:comparison} \\
\hline\hline
\multicolumn{3}{c}{} & \multicolumn{2}{c}{This work} & \multicolumn{2}{c}{Other theory}\\
\hline
{$Z$} & Resonance & {$J$} & {E} & {$\Gamma_{\rm Aug}$} & {E} & {$\Gamma_{\rm Aug}$} \\
\hline
\endfirsthead
\caption*{Table \ref{tab:comparison} ({\it Continued.})}\\\toprule
\multicolumn{3}{c}{} & \multicolumn{2}{c}{This work} & \multicolumn{2}{c}{Other theory}\\
\hline
{$Z$} & Resonance & {$J$} & {E} & {$\Gamma_{\rm Aug}$} & {E} & {$\Gamma_{\rm Aug}$} \\
\hline
\endhead
\hline
 5 &       $2s_{1/2}^2$ &  0 &   -5.662502(24) &       6.674(1)$\times 10^{-3}$ &        -5.66088\tnotex{a} & 6.650$\times 10^{-3}$ \tnotex{a} \\ 

   &       $2p_{3/2}^2$ &  0 &   -5.145465(42) &         3.0(2)$\times 10^{-4}$ &        -5.14461\tnotex{a} & 3.010$\times 10^{-4}$ \tnotex{a} \\ 

   & $2s_{1/2}2p_{1/2}$ &  0 &   -5.614844(13) &       3.314(7)$\times 10^{-4}$ &        -5.61299\tnotex{a} & 3.208$\times 10^{-4}$ \tnotex{a} \\ 

   &                    &  1 &   -5.614581(13) &       3.277(7)$\times 10^{-4}$ &                           &                                \\ 

   & $2s_{1/2}2p_{3/2}$ &  2 &   -5.613991(13) &       3.241(5)$\times 10^{-4}$ &                           &                                \\ 

 6 &       $2s_{1/2}^2$ &  0 &   -8.291878(40) &       6.944(1)$\times 10^{-3}$ &        -8.28820\tnotex{a} & 6.910$\times 10^{-3}$ \tnotex{a} \\ 

   &       $2p_{3/2}^2$ &  0 &   -7.653119(54) &         3.1(1)$\times 10^{-4}$ &        -7.65106\tnotex{a} & 3.210$\times 10^{-4}$ \tnotex{a} \\ 

   & $2s_{1/2}2p_{1/2}$ &  0 &   -8.234485(22) &       3.377(6)$\times 10^{-4}$ &        -8.23029\tnotex{a} & 3.220$\times 10^{-4}$ \tnotex{a} \\ 

   &                    &    &                 &                                &       -8.234485\tnotex{b} & 3.392$\times 10^{-4}$ \tnotex{b} \\ 

   &                    &  1 &   -8.233915(22) &        3.32(1)$\times 10^{-4}$ &       -8.233914\tnotex{b} & 3.327$\times 10^{-4}$ \tnotex{b} \\ 

   & $2s_{1/2}2p_{3/2}$ &  2 &   -8.232652(22) &        3.27(1)$\times 10^{-4}$ &       -8.232654\tnotex{b} & 3.269$\times 10^{-4}$ \tnotex{b} \\ 

 7 &       $2s_{1/2}^2$ &  0 &  -11.422672(64) &       7.146(1)$\times 10^{-3}$ &       -11.41546\tnotex{a} & 7.100$\times 10^{-3}$ \tnotex{a} \\ 

   &       $2p_{3/2}^2$ &  0 &  -10.661511(64) &         3.3(2)$\times 10^{-4}$ &       -10.65732\tnotex{a} & 3.340$\times 10^{-4}$ \tnotex{a} \\ 

   & $2s_{1/2}2p_{1/2}$ &  0 &  -11.355781(34) &       3.444(2)$\times 10^{-4}$ &       -11.34755\tnotex{a} & 3.230$\times 10^{-4}$ \tnotex{a} \\ 

   &                    &  1 &  -11.354691(34) &       3.363(6)$\times 10^{-4}$ &                           &                                \\ 

   & $2s_{1/2}2p_{3/2}$ &  2 &  -11.352301(34) &       3.289(5)$\times 10^{-4}$ &                           &                                \\ 

 8 &       $2s_{1/2}^2$ &  0 &  -15.055424(96) &       7.304(1)$\times 10^{-3}$ &       -15.04266\tnotex{a} & 7.250$\times 10^{-3}$ \tnotex{a} \\ 

   &       $2p_{3/2}^2$ &  0 &  -14.171026(76) &         3.4(2)$\times 10^{-4}$ &       -14.16345\tnotex{a} & 3.440$\times 10^{-4}$ \tnotex{a} \\ 

   & $2s_{1/2}2p_{1/2}$ &  0 &  -14.979340(50) &       3.516(5)$\times 10^{-4}$ &       -14.96481\tnotex{a} & 3.235$\times 10^{-4}$ \tnotex{a} \\ 

   &                    &  1 &  -14.977439(49) &       3.409(6)$\times 10^{-4}$ &                           &                                \\ 

   & $2s_{1/2}2p_{3/2}$ &  2 &  -14.973291(50) &       3.304(5)$\times 10^{-4}$ &                           &                                \\ 

 9 &       $2s_{1/2}^2$ &  0 &   -19.19081(14) &       7.434(3)$\times 10^{-3}$ &       -19.16983\tnotex{a} & 7.365$\times 10^{-3}$ \tnotex{a} \\ 

   &       $2p_{3/2}^2$ &  0 &  -18.182130(88) &         3.5(3)$\times 10^{-4}$ &       -18.16951\tnotex{a} & 3.520$\times 10^{-4}$ \tnotex{a} \\ 

   & $2s_{1/2}2p_{1/2}$ &  0 &  -19.105923(70) &       3.600(7)$\times 10^{-4}$ &       -19.08204\tnotex{a} & 3.240$\times 10^{-4}$ \tnotex{a} \\ 

   &                    &  1 &  -19.102832(70) &       3.463(7)$\times 10^{-4}$ &                           &                                \\ 

   & $2s_{1/2}2p_{3/2}$ &  2 &  -19.096092(70) &       3.326(8)$\times 10^{-4}$ &                           &                                \\ 

10 &       $2s_{1/2}^2$ &  0 &   -23.82944(19) &       7.542(1)$\times 10^{-3}$ &       -23.79699\tnotex{a} & 7.460$\times 10^{-3}$ \tnotex{a} \\ 

   &       $2p_{3/2}^2$ &  0 &   -22.69519(10) &         3.6(2)$\times 10^{-4}$ &       -22.67551\tnotex{a} & 3.585$\times 10^{-4}$ \tnotex{a} \\ 

   & $2s_{1/2}2p_{1/2}$ &  0 &  -23.736271(96) &       3.690(4)$\times 10^{-4}$ &       -23.69927\tnotex{a} & 3.243$\times 10^{-4}$ \tnotex{a} \\ 

   &                    &  1 &  -23.731510(96) &       3.518(3)$\times 10^{-4}$ &                           &                                \\ 

   & $2s_{1/2}2p_{3/2}$ &  2 &  -23.721110(96) &       3.346(7)$\times 10^{-4}$ &                           &                                \\

\hline\hline 
\end{longtable}
\begin{tablenotes}
  \footnotesize
  \item[a] \label{a} Ho~\cite{Ho_PRA23_2137:1981}
  \item[b] \label{b} M\"uller {\it et al.}~\cite{Muller_PRA98_033416:2018}
\end{tablenotes}
\end{ThreePartTable}
\end{center}
\endgroup
%
%
%
\section{CONCLUSION}
The energies and Auger widths of the $LL$ resonances of the He-like ions from boron ($Z = 5$) to argon ($Z = 18$) have been      evaluated by means of the complex scaled configuration-interaction method.
The systematic analysis of the uncertainty arising from the limited size of the configuration space was performed.
The obtained energies have been compared with the ones calculated using the stabilization and basic balancing methods. 
It was found that the energies obtained with these methods differ from the complex scaling results by a shift that varies from about 
$1$~meV to $10$~meV.
%
%
\\ \indent
%
%
The nuclear recoil and QED corrections were evaluated separately and added to the complex rotated Dirac-Coulomb-Breit energies.
As the result, the most accurate theoretical predictions for the energies of the $LL$ resonances are obtained.
In most cases, the accuracy of the total results is limited by 
the uncertainties from the higher-order QED corrections.
%
%
\\ \indent
%
%
%
\section{ACKNOWLEDGMENTS}
This work was supported by the Russian Science Foundation (Grant No. 17-12-01097).
%
%
%

%
\end{document}